\documentstyle[11pt]{article}

\newcommand{\kms}{\mbox{$\rm km\, s^{-1}$\,}}
\newcommand{\etal}{\mbox{\it et al.}\,}

\title{HI Observations of two Molecular Clouds with Extremely Large
       Velocity Dispersions}
\author{H. Riffert \\
       {\small Institut f\"ur Theoretische Astrophysik,
       Universit\"at T\"ubingen,} \\
       {\small Auf der Morgenstelle 10, D-72076 T\"ubingen, Germany}
       \and 
       P. Kumar\thanks{Alfred P. Sloan Fellow \& NSF Young Investigator} \\
       {\small Dept. of Physics, MIT, Cambridge, MA 02139} 
       \and
       W.K. Huchtmeier\\
       {\small Max Planck Institut f\"ur Radioastronomie,}\\ 
       {\small Auf dem H\"ugel 69, D-53121 Bonn, Germany}}
\date{}

\begin{document}
\maketitle

\newpage
\begin{abstract}
\noindent
We have mapped two molecular clouds at ($\ell,\, b$) = ($3.2^\circ, + 
0.3^\circ$) and ($\ell,\, b$) = ($5.4^\circ,\, -0.5^\circ$) in 21-cm 
line and continuum emission. These clouds show unusually large velocity 
dispersions of more than 100~\kms (FWHM) which has also been seen in 
$^{12}$CO, $\rm ^{13}CO$, and CS emissions. This dispersion is roughly an 
order of magnitude larger compared to giant molecular clouds. From our HI 
observation we estimate that the atomic mass of the cloud at $\ell = 
3.2^\circ$ is about $1.5\times 10^5\,{\rm M_\odot}$ and the mass of the 
cloud at $\ell = 5.4^\circ$ is $7\times 10^4\,{\rm M_\odot}$. The ratio of 
total molecular mass to atomic mass for these clouds appears to be normal 
for clouds near the galactic center. The main peculiar feature of these 
clouds is their abnormally large velocity dispersion; the extent in 
velocity is from about 0 km s$^{-1}$ (LSR) to 200 km s$^{-1}$. These clouds
are connected, in the $\ell$--v plane, to high velocity ridges that extend
over several degrees of the longitude. These properties, we believe, 
provide important clues to the physical process responsible for their large
velocity dispersion.

\end{abstract}

\clearpage

\section{Introduction}
There are more than 100 giant molecular clouds in the Galaxy with mass of
10$^6~$M$_\odot$ or greater. The internal velocity dispersion of these clouds,
measured using $^{12}$CO emission, is in most cases approximately equal to 
their virial velocity of about 15~\kms (Dame \etal 1986).
However, there are at least two very unusual molecular clouds,
located at $(\ell,b)=$($3.2^\circ,\, 0.3^\circ$) and ($5.4^\circ,\, 
-0.5^\circ$), which have extremely large velocity dispersions
of about 100~\kms (FWHM) as seen in several molecular lines including
$^{12}$CO. Hereafter we shall refer to these objects as wide 
line clouds (WLCs). Oort drew attention
to the large velocity dispersion of the WLC at $3.2^\circ$ longitude
in his influential review article almost two decades ago (1977).
However, these clouds remained largely unexplored until the mid-eighties
when Stark and Bania (1986) and Bitran (1987) mapped them
at high spatial resolution in $^{12}$CO, $^{13}$CO, and CS, and confirmed
their unusually large velocity dispersion. The observed velocity dispersion
even in CS, which is associated with high density cores of clouds, is
about $40$~\kms compared to $5$--$10$~\kms for giant
molecular clouds.
Both clouds are also visible in the large-scale OH absorption-line survey of
Boyce and Cohen (1994a) where the properties of seven WLCs are discussed in
the longitude range $354^{\circ} < \ell < 6^{\circ}$. For an overall
$\ell$--$b$ map of this galactic centre region see Boyce and Cohen (1994b).
In addition, the same authors have mapped the cloud at $\ell = 5.4^{\circ}$
using $\rm H_2CO$ measurements and give a mass estimate of $5\times 10^6\, {\rm M_\odot}$
(Boyce {\it et al.} 1989).

According to Bitran (1987) and Stark and Bania (1986) the mass of the clouds, 
estimated from the $^{12}$CO emission, is about $10^7~M_\odot$, however,
this is very uncertain because the ratio of H$_2$ mass to CO emission
for clouds in the inner few kpc is likely to be different from the 
solar neighborhood (Stacy \etal 1988).
We have mapped the two WLCs in $21$-cm in order to provide 
a lower limit to their mass, to determine their 
velocity dispersion, and to ascertain if they are different from
giant molecular clouds. 

In the next section we describe the observational details, Sect.~3
contains the results and a comparison with the CO-observations.
The main results are summarized in Sect.~4.

\section{Observations}
We have observed the two wide-line clouds in $21$-cm with the Effelsberg
$100$~m telescope using a cooled HEMT receiver with a system temperature 
less than about $30$~K on cold sky but increased to about $55$~K due to radio
emission from the galactic centre and spillover from the ground. 
The total bandwith of the receiver was $3.12$~MHz around the neutral hydrogen
hyperfine transition frequency of $1420.4$~MHz; the
corresponding velocity range was $-229.7$ to $429.3$~\kms.
Using the $1024$ channel autocorrelation spectrometer
a resolution of about $1.3$~\kms was obtained after
Hanning filtering of the spectra. The clouds have been
mapped in steps of $4.5^\prime$ which is half the beam width of the
telescope. The total angular coverage in $\ell$ and $b$ for the two
clouds were ($2.675^\circ , \, -0.450^\circ$)--($3.725^\circ , \, 0.975^\circ$)
and ($4.875^\circ , \, -1.250^\circ$)--($5.925^\circ , \, 0.175^\circ$),
respectively.
Thus each cloud was scanned at $300$ observation points with
an integration time of $15$~s per position.
The temperature was calibrated by periodically switching to 
a stabilized noise tube, and the calibration of the noise tube 
was done by observing the standard area S7 before and after
mapping each cloud. 
The resulting temperature calibration factors
are used to convert the measured voltage to brightness temperature
including atmospheric extinction. For both clouds the rms noise level
in the raw data was about $700$~mK.
We did not however correct the maps for effects from the antenna side lobes;
the resulting error is expected to be small (the relative sensitivity of the
first side lobes, separated from the main beam by $14^\prime$, is about $1\%$).

At low velocities the observed spectra are dominated by strong emission from 
foreground material. The WLCs can be seen as broad emission structures
extending up to $V_{LSR} \approx 200$~\kms.
At both ends of the observed velocity interval
no neutral hydrogen emission was detected, and we have 
used these emission free regions to subtract a linear baseline
from each spectrum with a baseline error of about $0.3$~K.

\section{Results and Discussion}
In Fig.~1 we show average $21$-cm spectra of the clouds, i.e. the brightness
temperature averaged over the observed area. 
For the WLC at $\ell = 3.2^\circ$ this
average contains three distinct peaks. The first one at
$V_{LSR} \approx 90$~\kms is caused by the emission from the WLC
itself, the second one at $V_{LSR} \approx 165$~\kms is partly from the
WLC and also contains a strong contribution from an arm-like
structure at $b \approx 0.5$ which extends over the entire observed
$\ell$-range (compare Fig.~2). The high velocity peak at $V_{LSR} \approx
220$~\kms is entirely due to HI near the galactic center unrelated to the
WLC.

In the composite HI spectra of the WLC at $\ell \approx 5.4^\circ$ (fig. 1)
two peaks can be seen. The one at $V_{LSR} \approx 85$~\kms is probably
unrelated to the WLC. The strong peak at $V_{LSR} \approx 185$~\kms
is again due to a extended arm-like structure which is however
more uniformly distributed over the observed area with a strong 
contribution from emission at $b \approx -1^\circ$.
%{\bf Harald: looking at fig. 3 it appears that b=-0.2}.
This WLC is weaker than the WLC at $\ell = 3.2^\circ$ 
and is harder to recognize in the average spectrum.

Fig.~2 shows $\ell$-$v$-maps for both of the observed regions. 
The WLCs are clearly seen and appear to 
cover a large velocity range of more than $100$~\kms.
It is a remarkable fact that in both cases the clouds end at
high velocity ridges extending over the entire observed longitude range.
%
%The emission from the ridge that appears to be associated with the
%$3.2^\circ$ degree cloud comes primarily from $\ell>3^\circ$
%(i.e. the ridge is to the left of the cloud), whereas the ridge for the 
%WLC at $\ell = 5.4^\circ$ is nearly symmetrical about the cloud.
%
These ridges can also be identified in the corresponding 
CO-maps, although the ratio of intensities between the clouds and 
the ridges are much larger in CO compared to HI, i.e. the ridges appear
much weaker in CO.

Contour maps of the temperature integrated over the velocity range 
from $50$~\kms to $200$~\kms as a function of $\ell$ and $b$ are shown 
in Fig.~3 for both WLCs in 21-cm as well as $^{12}$CO emission.
Both WLCs clearly stand out against the background, and
their positions, sizes, and orientations obtained in 
HI coincide with the $\rm ^{12}CO$ emission; it is evident 
that we are looking at the same objects in both of these lines.
There is a small offset of about $2$~arcmin between the centers
of the HI- and the $\rm ^{12}CO$-map for the WLC at $\ell=3.2^\circ$.
This is however insignificant since the shift is much less than
the beam width for both observations.
It can also be seen that the WLC at $\ell = 5.4^\circ$ is partly
obscured by the HI-emission from the galactic plane but stands out
more strongly in $\rm ^{12}CO$.

The WLCs are almost certainly not nearby objects. The reason for this 
is that the observed velocity range of the WLCs seen in $21$-cm, 
as well as $^{12}$CO, is approximately 0--200 km s$^{-1}$, with 
a mean velocity of about 100 km s$^{-1}$. Thus if these clouds were 
nearby objects on roughly circular orbit, then their orbital velocity 
projected along the line of sight would be small and therefore their 
observed internal random velocity should have comparable negative and 
positive values. This is contrary to the observations. The observed 
velocity of an object, on a circular orbit, in the galactic plane at 
longitude $\ell$ is
\begin{displaymath}
V_{LSR} = r_\odot[\Omega(r)-\Omega(r_\odot)]\sin\ell
\end{displaymath}
where $r_\odot$ and $r$ are
the distances of the observer and the object from the galactic center, 
respectively, and $\Omega$ is the angular rotation speed. 
Taking $r\Omega$ to be approximately constant we find the 
distance of the WLC at $\ell=3.2^\circ$ from the galactic center 
to be $r_\odot/10$, and the one at $\ell=5.4^\circ$ to be $r_\odot/6.4$. 
Thus the distance $d$ of the WLCs from us is approximately $r_\odot$.

The total neutral hydrogen mass of the WLCs is estimated from the observed
flux in $21$-cm using the formula 

\begin{equation}
\frac{M_{\rm HI}}{M_\odot} = 3.213 \times 10^{2} 
\left( \frac{d}{8.5~{\rm kpc}} \right)^2
\int d\ell\,db\,dv\; T(\ell,b,v)~,
\end{equation}
where $T$ is brightness temperature in K,
$v$ is the velocity in \kms, and $\ell$ and $b$ are measured in degrees.
This expression follows from the number of hydrogen atoms along the line
of sight as given by Burton (1988) multiplied with the atomic hydrogen mass
and the suface element $dA = d^2\, \cos{b}\, d\ell\, db \approx 
d^2 \, d\ell\, db$.

Substituting the observed quantities in the above equation we find 
that the HI mass in the cloud regions, i.e. without subtracting any
background emission, are about $3\times 10^5~{\rm M_\odot}$
for the $\ell = 3.2^\circ$ cloud, and $10^5~{\rm M_\odot}$
for the cloud at $\ell = 5.4^\circ$. These values were obtained by 
integrating the brightness temperature over the velocity range of 
50--200~$\rm km~s^{-1}$ in order to eliminate most of emission from 
foreground HI. However, there is a substantial $21$-cm emission 
in this velocity range associated with the galactic plane at 
the angular location of both of the WLCs
that must also be subtracted in order to determine the atomic mass of
these clouds. We have tried several different
ways of estimating this {\it background} emission. One method 
is to calculate the total HI emission outside the 
region occupied by each cloud, and multiply it by the fractional 
area covered by the cloud to yield the contribution of the 
background emission. This procedure gives the {\it corrected} HI mass
of the  WLC at $3.2^\circ$ to be $8\times 10^4~{\rm M_\odot}$, and 
the WLC at $5.4^\circ$ to be about $3\times 10^4~{\rm M_\odot}$.
However, this method overestimates the {\it background} contribution,
or underestimates cloud mass,
since the emission from outside the cloud region is very nonuniform and is
dominated by the strong $21$-cm emission from the galactic plane.
We have also subtracted average spectra from the cloud region 
before calculating the mass from the integral of equation (1); 
various areas outside the clouds were used to determine the average 
spectra. When the galactic plane is not included in the area that
is used in estimating the average spectra, 
then we obtain cloud mass that are approximately the averages of the
two extreme values mentioned above, and is likely to be closer to the
actual atomic mass of these clouds which lie outside of the
galactic plane. Our best estimate of the
HI mass of the WLC at $3.2^\circ$ is $1.5\times 10^5 M_\odot$, and the 
WLC at $5.4^\circ$ is $7\times 10^4 M_\odot$ with an uncertainty 
of about a factor of 2. 

The HI mass is obviously the lower limit to the total cloud mass, most
of which is molecular hydrogen.
Neutral hydrogen in a typical giant molecular clouds is observed in a
halo surrounding the cloud, with a ratio of HI to molecular mass of
roughly 10\% (e.g. Anderson, Wannier \& Morris 1991, Elmegreen \& Elmegreen
1987). The neutral hydrogen mass of the WLCs is about 1\%
of their total mass as determined from the $^{12}$CO emission
(Bitran 1987). It has been suggested that the ratio of H$_2$ mass to
$^{12}$CO emission for clouds near the
galactic center is perhaps smaller by a factor of about 10 compared
to the value in the solar neighborhood (see Stacy \etal 1988). 
We also expect to see a larger fraction of the $^{12}$CO molecules in these
clouds because of their large velocity dispersion. Taking these factors
into consideration we conclude that the ratio of the HI mass to the 
total mass for the WLCs is not very different from giant 
molecular clouds.

\section{Summary}
We have mapped two molecular clouds at $\ell = 3.2^\circ$ and
$\ell = 5.4^\circ$ in the $21$-cm line emission of neutral hydrogen.
These clouds show unusually large velocity dispersions of more than
$100$~\kms (FWHM) which has also been seen in lines of $\rm ^{12}CO$, 
$\rm ^{13}CO$, and $\rm CS$. The position, size, and orientation of both 
of the wide line clouds (WLCs) are the same in HI and $^{12}$CO maps. 
The WLCs are very prominent features in the $\ell$-$v$-maps and show up 
as highly extended vertical bands.

These clouds lie about 0.3$^\circ$ from the Galactic plane and are almost 
certainly within about 1 kpc of the Galactic center. 
The mass of the clouds calculated from their $\rm ^{12}CO$
emission, using the standard ratio of H$_2$ to $^{12}$CO mass,
is about $10^7~{\rm M_\odot}$ (Bitran 1987); however for clouds in 
the Galactic center region the ratio of H$_2$ mass to $^{12}$CO luminosity
is perhaps smaller by a 
factor of about 10 compared to the value in the solar neighborhood. From 
our HI observation we estimate the atomic mass of the WLCs at 
$\ell = 3.2^\circ$ and $5.4^\circ$ to be $1.5\times 10^5\,{\rm M_\odot}$, and  
$7\times 10^4\,{\rm M_\odot}$ respectively, with an
uncertainty of about a factor of $2$ due to the 
error in the determination of the background emission.
The ratio of HI to molecular mass of the WLCs does not appear to be
abnormal for clouds near the Galactic center, provided we use the 
ratio of H$_2$ to $^{12}$CO preferred for this region.

It is striking that the WLCs extend in velocity from roughly 0 km s$^{-1}$
(LSR) to about 200 km s$^{-1}$. Moreover, both of these clouds 
appear to end at high velocity ridges, in the $\ell$--v plane, that 
are very bright in 21-cm and extend over the entire observed longitude 
range (Fig.~2). 

These ridges are also visible in 
$^{12}$CO maps extending to several degrees in longitude; however the 
ridges are much less prominent in CO compared to HI.
Boyce and Cohen (1994a) have found similar structures (which they call
filaments) in their OH absorption-line measurements, and these filaments
seem to connect several of the seven WLCs described in this survey.
We believe that these features provide important constraints 
to the physical origin of the enormous velocity dispersion of the WLCs.

\medskip
\noindent {\bf Acknowledgment:} PK is grateful to Pat Thaddeus for suggesting
this problem, numerous discussions and the use of his CO data. 
We are indebted to Tom Dame for many interesting discussions about
molecular clouds and for suggesting several improvements to the presentation
of our work. This work was carried out while PK was visiting the Institute 
for Advanced Study, Princeton. He is very grateful to John Bahcall for his 
comments, support and hospitality.

\clearpage
\section*{References}
\begin{description}
  \item Anderson, B.G., Wannier, P.G., Morris, M. 1991,
   {\it Astrophys. J.}, {\bf 366}, 464 

  \item Bitran, M.E. 1987,
   {\it Ph.D. Thesis}, University of Florida

  \item Boyce, P.J., Cohen, R.J., Dent, W.R.F. 1989,
   {\it Mon. Not. R. astr. Soc.}, {\bf 239}, 1013

  \item Boyce, P.J., Cohen, R.J. 1994a,
   {\it Astron. Astrophys. Suppl. Ser.}, {\bf 107}, 563

  \item Boyce, P.J., Cohen, R.J. 1994b,
   {\it IAU Symp.} {\bf 169}, 311, eds. L. Blitz \& P. Teuben, Kluwer press 

  \item Burton, W.B 1988,
  in: {\it Galactic and Extragalactic Radio Astronomy}\, ($2^{\rm nd}$ ed.), 
  Eds.: G.L. Verschuur \& K.I. Kellermann, p. 295,
  Springer-Verlag, Berlin. 

  \item Dame, T.M., Elmegreen, B.G., Cohen, R.S., Thaddeus, P., 1986, 
  {\it Astrophys. J.}, {\bf 305}, 892

  \item Dame, T.M., Ungerechts, H., Cohen, R.S., de~Geus, E.J.,
   Grenier, I.A., May, J., Murphy, D.C., Nyman, L.-\AA., Thaddeus, P. 1987,
   {\it Astrophys. J.}, {\bf 322}, 706 

  \item Elmegreen, B.G., \& Elmegreen D.M., 1987, ApJ 320, 182

  \item Oort, J. 1977,
   {\it Ann. Rev. Astron. Astrophys.}, {\bf 15}, 295 

  \item Stacy, J.G., Bitran M.E., Dame, T.M., Thaddeus, P. 1988, 
   {\it The Galactic Center (IAU Symp.)}~{\bf 136}), 157, ed. M. Morris, 
   Reidel, Dordrecht

  \item Stark A.A, Bania T.M. 1986,
   {\it Astrophys. J.}, {\bf 306}, L17

\end{description}

\newpage
\section*{Figure Captions}
\noindent{\bf Fig.~1.} --
           Composite HI spectra of the WLCs as a function
           of velocity. The solid lines show the brightness temperature 
           averaged over the entire observed area. The dashed lines are the
           corresponding variances. The left/right panels are for clouds
           at $\ell=3.2^\circ/5.4^\circ$

\bigskip
\noindent{\bf Fig.~2.} --
           Longitude-velocity contour maps of both WLCs.
           The maps show the brightness temperature integrated over
           the $b$-range (\,$0.15^\circ,\, 1.0^\circ$) for
           the cloud at $\ell = 3.2^\circ$, and over
           ($-1.0^\circ,\, -0.15^\circ$) for the
           $\ell=5.4^\circ$ cloud.
           Contour levels are separated by $0.5$~{\rm K\, deg}.

\bigskip
\noindent{\bf Fig.~3.} --
           Longitude-latitude contour maps of both WLCs.
           The solid lines represent the HI data (brightness temperature), and
           the dashed lines show the $^{12}$CO data (antenna temperature).
           Velocity integration ranges are (\,$50,~200$)~\kms
           and (\,$40,~240$)~\kms for the HI and $\rm ^{12}CO$
           maps, respectively. No background subtraction has been
           performed for these maps.
           The corresponding contour levels are separated by
           $75$ {\rm K} \kms (HI) and $50$ {\rm K} \kms ($\rm ^{12}CO$) for
           the WLC at $\ell = 3.2^\circ$, and $50$ {\rm K} \kms (HI) and
           $20$ {\rm K} \kms ($\rm ^{12}CO$) for the $\ell = 5.4^\circ$ cloud.
	   The $^{12}$CO data is courtesy of P. Thaddeus \& T. Dame, and was
           obtained using a 1.2 meter telescope; for details please
           see Dame et al. (1987) \& Bitran (1987).

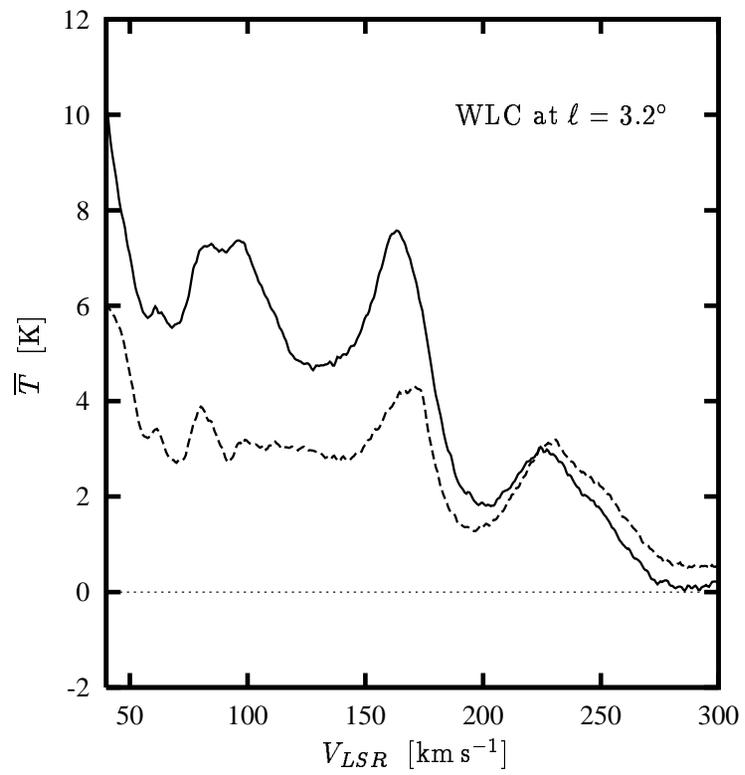
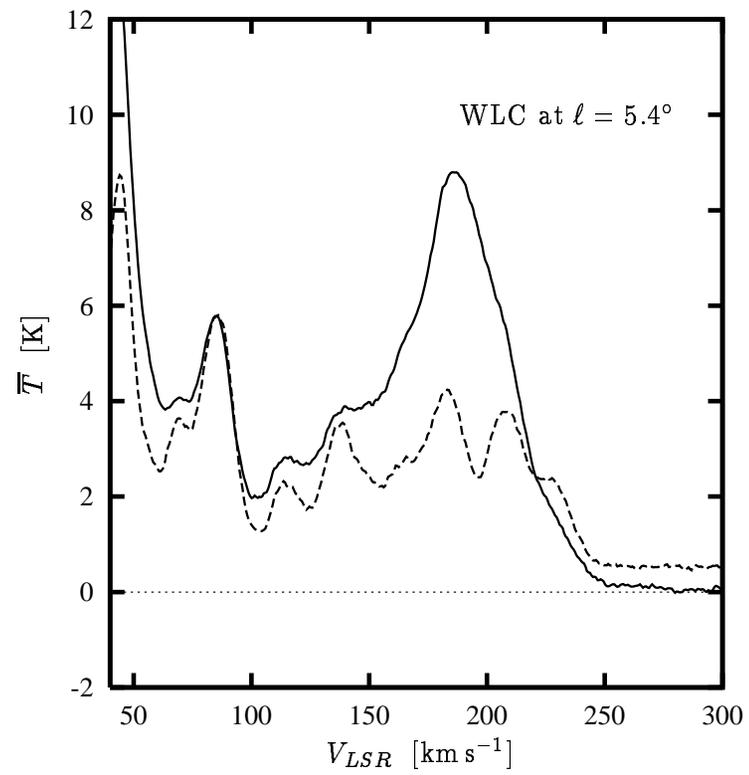
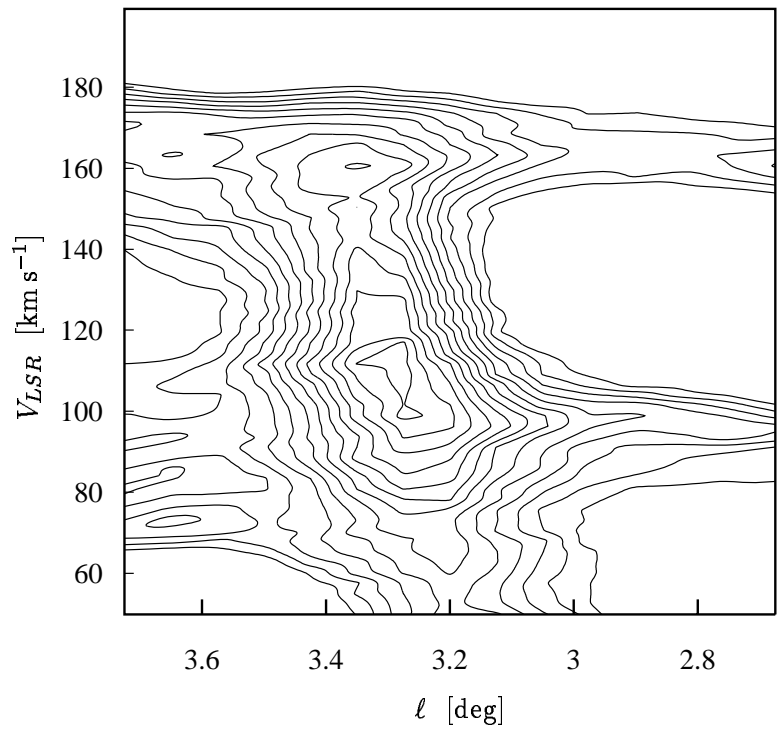
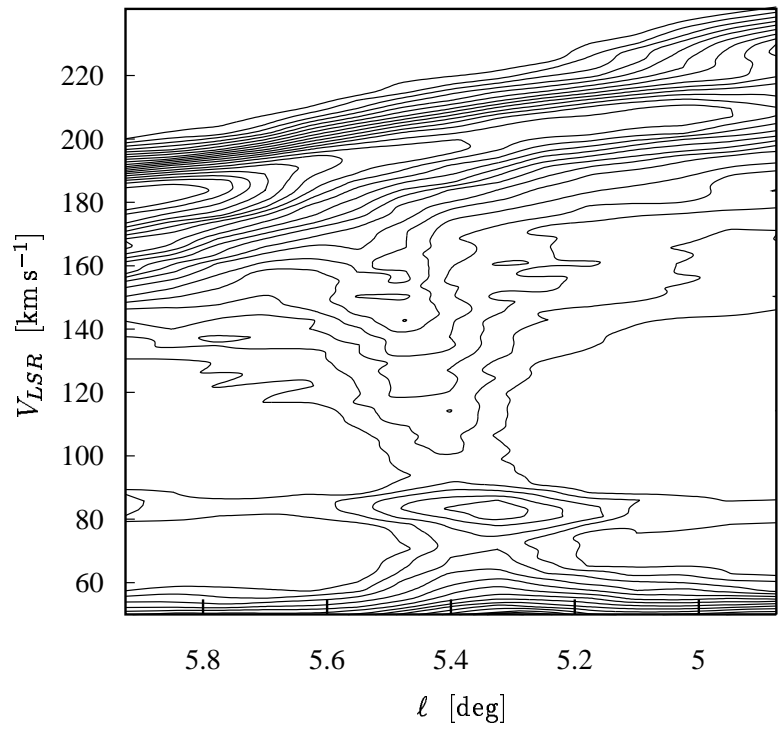
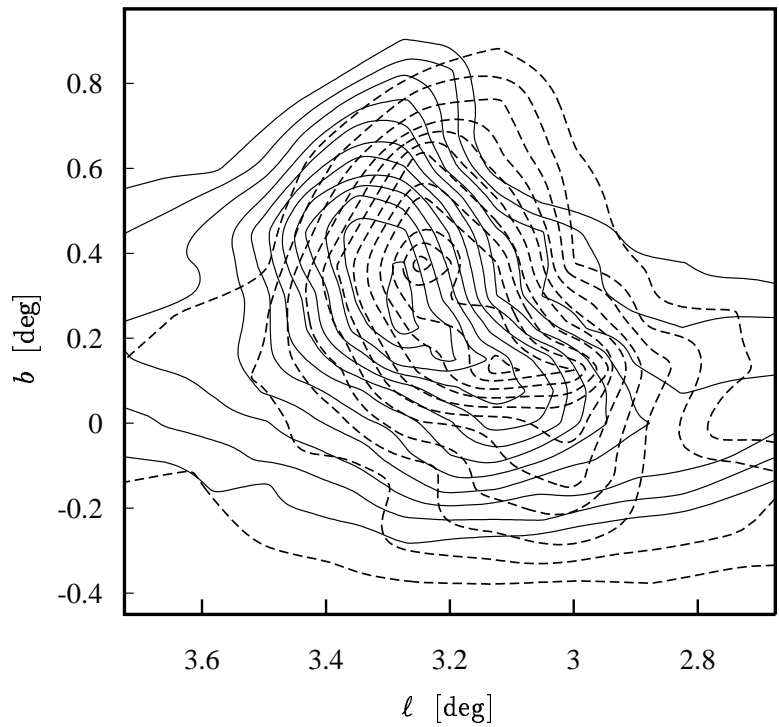
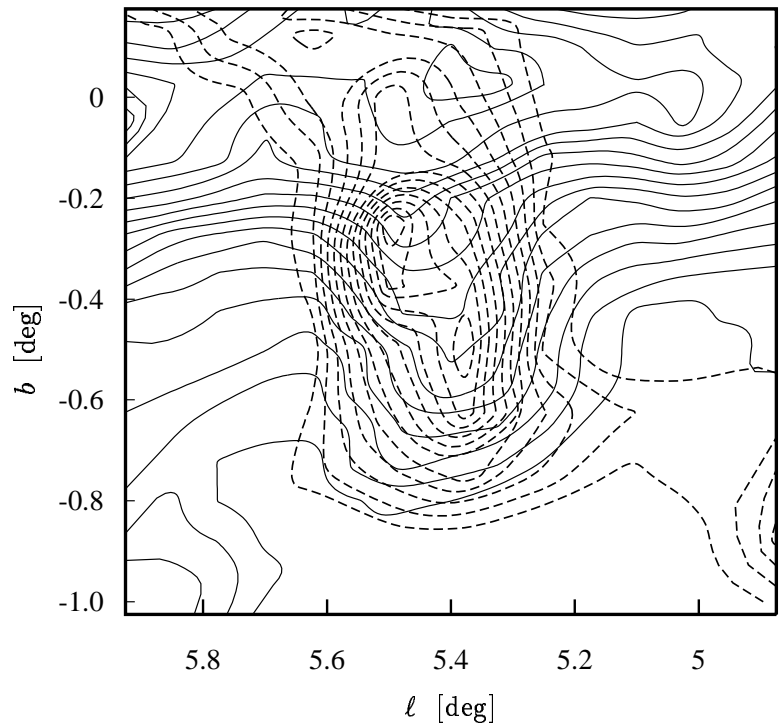
\end{document}